\documentclass[
]{ceurart}

\usepackage{listings}
\usepackage{amsmath,amssymb,amsfonts}
\usepackage{algorithmic}
\usepackage{graphicx}
\usepackage{textcomp}
\usepackage{xcolor}
\usepackage[caption=false]{subfig}
\usepackage[many,minted]{tcolorbox}
\usepackage{caption}

\usepackage{hyperref}

\newcommand{\mynewminted}[3]{%
  \newminted[#1]{#2}{#3}%
  \tcbset{myminted/#1/.style={minted language=#2,minted options={#3}}}
}

\mynewminted{myc}{c}{tabsize=2,fontsize=\footnotesize,linenos, numbersep=3mm,escapeinside=||}

\newtcblisting[auto counter,number within=section,
  list inside=mypyg]{listingsbox}[3][]{%
  listing only,
  title={Listing \thetcbcounter: #3},
  list entry={\protect\numberline{\thetcbcounter}#3},
  enhanced,
  left=6mm,
  overlay={\begin{tcbclipinterior}\fill[black!15] (frame.south west)
            rectangle ([xshift=6mm]frame.north west);\end{tcbclipinterior}},
  colframe=black!35,
  drop fuzzy shadow,
  myminted/#2,#1
}

\begin{document}

\copyrightyear{2025}
\copyrightclause{Copyright for this paper by its authors.
  Use permitted under Creative Commons License Attribution 4.0
  International (CC BY 4.0).}

\conference{GeCoIn 2025: Generative Code Intelligence Workshop, co-located with the 28th European Conference on Artificial Intelligence (ECAI-2025),  October 26, 2025 --- Bologna, Italy}

\title{UnitTenX: Generating Tests for Legacy Packages with AI Agents Powered by Formal Verification}

\author[1]{Yiannis Charalambous}[%
orcid=0009-0000-5755-5099,
email=yiannis.charalambous-4@postgrad.manchester.ac.uk,
url=https://yiannis.site,
]
\cormark[1]
\address[1]{The University of Manchester, UK}

\author[2]{Claudionor N. Coelho Jr}[%
orcid=0000-0001-9637-1890,
email=claudionor.coelho@alumni.stanford.edu,
url=https://www.linkedin.com/in/claudionor-coelho-jr-b156b01/,
]
\address[2]{ECE Department, Santa Clara University, US}

\author[3]{Luis Lamb}[%
orcid=000-0003-1571-165X,
email=lislamb@acm.org,
url=linkedin.com/in/luis-lamb-131394,
]
\address[3]{UFRGS, Brazil}

\author[1,4]{Lucas C. Cordeiro}[%
orcid=0000-0002-6235-4272,
email=lucas.cordeiro@manchester.ac.uk,
url=https://ssvlab.github.io/lucasccordeiro/,
]
\address[4]{UFAM, Brazil}

\cortext[1]{Corresponding author.}

\begin{abstract}
This paper introduces \textit{UnitTenX}, a state-of-the-art open-source AI multi-agent system designed to generate unit tests for legacy code, enhancing test coverage and critical value testing. \textit{UnitTenX} leverages a combination of AI agents, formal methods, and Large Language Models (LLMs) to automate test generation, addressing the challenges posed by complex and legacy codebases.
Despite the limitations of LLMs in bug detection, \textit{UnitTenX} offers a robust framework for improving software reliability and maintainability. Our results demonstrate the effectiveness of this approach in generating high-quality tests and identifying potential issues. Additionally, our approach enhances the readability and documentation of legacy code.
\end{abstract}

\begin{keywords}
    Artificial Intelligence \sep
    Large Language Models \sep
    Formal Methods \sep
    Software Verification \sep
    Software Engineering
\end{keywords}

\maketitle

\section{Introduction}
\label{sec:intro}

Software testing is an important element of the software engineering life-cycle that ensures the software is correct~\cite{ahamed2010studyingfeasibilityimportancesoftware}. In the past, writing tests for software was neglected as it was not deemed as important as it is in recent times~\cite{felderer2020evolution}.
Legacy software is defined as software that uses outdated technologies with source code that is not actively maintained, but is actively used in production~\cite{IBM2025LegacyCode}. According to~\cite{feathers2004working}, legacy code may contain little or no tests.
This neglect is oftentimes reflected in modern software due to deadline restrictions~\cite{marques2017survey}. In many cases, this lack of tests can manifest as bugs and security vulnerabilities in legacy and newly built software~\cite{esther_managing_2024, smyth2023penetrationtestinglegacysystems}. Contemporary examples of high-profile bugs in complex modern systems are the Heartbleed bug~\cite{banks2015heartbleed} and the CrowdStrike~\cite{por2024systematic} outage. These incidents underscore the importance of thoroughly testing legacy codebases and ensuring the correctness of software systems. 

For example, consider a sequence of $1000$ independent and uncorrelated \textit{if-then-else} statements, such as those found in network-based policy devices~\cite{hansen_abhdjbdns_2025}; this scenario alone presents around $2^{1000}$ possible states to verify. Legacy code is often highly complex and lacks documentation, which makes bug detection challenging. According to~\cite{tornhill2022code}, such conditions can result in up to $15$ times more defects and extend the time required to develop new features by as much as $124$\%. The increase of bugs is reflected by the reduced predictability inherent in poorly maintained codebases.

To address this problem, we propose the use of AI Agents, which are software entities that have access to a limited environment where they can perform actions that change the environment autonomously~\cite{durante_agent_2024}. In the software testing field, AI code agents can be used to automate test creation by overseeing the execution of code to detect patterns, predict defects, and optimize test cases based on software requirements and past results~\cite{huang2023agentcoder}. This auto-regressive stochastic behavior allows them to improve the accuracy and efficiency of the tests over time. The simplicity of use, efficiency, stability, and scalability make AI agents a powerful tool in software testing.

In this paper, we introduce \textit{UnitTenX}, an AI Agent that uses formal verification to identify and create unit tests for documenting the interfaces of legacy code, thereby uncovering conditions that can cause the software to crash. By extension, the unit tests created also act as regression tests since they are designed to test the software and achieve max coverage. This paper aims to answer the following Research Questions (RQs):

\begin{enumerate}
    \item How effectively does \textit{UnitTenX} generate unit tests that increase code coverage for legacy C codebases?
    \item How does \textit{UnitTenX} handle compilation errors, runtime exceptions (segmentation faults), and timeouts during automated test generation?
    \item How does the reflection and feedback loop in \textit{UnitTenX} contribute to iterative improvement of generated test suites?
\end{enumerate}

This paper advances the field through the following contributions:

\begin{itemize}
    \item Introduces UnitTenX, a formal verification driven AI agent tool that automatically generates unit tests for code-bases with no tests, addressing a critical challenge in maintaining and modernizing complex software systems such as legacy software.
    \item Combines symbolic analysis from tools like ESBMC with large language models to identify edge cases, crash conditions, and maximize code coverage in legacy C modules.
    \item Automates the creation of code mockups, enabling end-to-end test suite generation and integration with existing legacy infrastructure.
    \item Implements a reflection and feedback loop where a language model evaluates test outcomes, recommends improvements, and iteratively increases both the quality and coverage of generated test suites.
    \item Demonstrates robustness by recovering from common errors such as compilation faults and segmentation violations, resulting in production-ready regression suites on real-world legacy software.
\end{itemize}

This paper is organized as follows: In Section~\ref{sec:preliminaries} we describe the background theory that \textit{UnitTenX} uses and also work analogous to \textit{UnitTenX} and distinguish the areas in which it innovates. in Section~\ref{sec:motivation}, we present a motivating example for using \textit{UnitTenX}, and discuss the limitations of AI in bug detection, highlighting the challenges LLMs face in identifying bugs. In Section~\ref{sec:methodology}, we describe \textit{UnitTenX} itself and how it generates unit tests for legacy code. Section~\ref{sec:eval} presents the results of our study, demonstrating the effectiveness of our approach. Finally, Section~\ref{sec:conclusion} discusses threats to validity and concludes the paper, discussing the limitations of our approach and potential areas for future research.

\section{Preliminaries}
\label{sec:preliminaries}

The following section covers the intersection of the theories that \textit{UnitTenX} utilizes, including unit test generation (the core background theory), formal methods (used to identify crash states in the legacy codebase), and large language models (used as the code generation element of \textit{UnitTenX}). Lastly, previous related works that are in the field are covered.

\subsection{Regression Testing}

Regression Testing is the process of testing software to detect any behavioral changes that may have appeared when the program is modified~\cite{korel1998automated, ramirez2023taxonomy}. This is usually done by running tests and checking if the Program Under Test (PUT) in each test has changed, this is used to verify that changes haven't introduced new bugs or altered existing functionality~\cite{630875}. Regression testing research can be split into Regression Test Selection and Optimization research~\cite{630875} and regression test generation research~\cite{verma2023software, guan2024review, arora2018systematic}. 
The main purpose of Regression Test Selection and Optimization research is to maximize the subset of test cases affected by the code changes. This is usually done using various static and dynamic analysis techniques. The selection process is primarily guided by code coverage, to maximize coverage and fault detection while minimizing cost and execution redundancy ~\cite{630875}.
In Regression Test Generation research, the primary purpose is to generate new test cases to test parts of the program that are not currently being tested. This is done through constraint solving at branches, using heuristics, templates, or using AI-based methods (which includes LLMs in recent years)~\cite{verma2023software, guan2024review, arora2018systematic}.

\subsection{Formal Verification}

Formal Verification is the process of encoding a program into an abstract representation to verify that it does not violate any predefined properties. 
In our experiments, we use the Efficient SMT-based Bounded Model Checker (ESBMC) to identify invalid states (through counterexamples) and generate test cases for the legacy software. ESBMC is a Bounded Model Checker (BMC) that encodes the PUT into an SMT formula and uses an SMT backend to find any violated program states.
As detailed in~\cite{monteiro2022model, biere2009handbook, barreto2011verifying}, ESBMC converts the program into a Control-Flow Graph (CFG), then extracts a state transition system $M=(S,T,s_0)$ where $S$ is the set of states and $T$ represents the transitions between the states $T \subseteq S \times S$ and $s_0$ represents the initial state of $M$. Let $I$ be a predicate evaluating the set of initial states of $M$. Given the state transition system $M$, a bound $k$, and a property $\phi$, the BMC process unrolls the system $k$ times and translates it into a Verification Condition (VC) $\psi$, where $\psi$ is satisfiable iff $\phi$ has a counterexample (CE) of length less than or equal to $k$. More formally, $I(s_0) \land \bigwedge_{i=0}^{k-1}T(s_i,s_{i+1})$ is the executions of $M$ of length $j$ and the formula can be satisfied iff there exists a state at step $j$ where $\phi$ is violated. The VC is calculated from the following formula:

\begin{equation}
\psi_k=I(s_0) \land \bigwedge_{i=0}^{k-1}T(s_i,s_{i+1})\land \bigvee_{i=0}^{k}\neg\phi(s_i)
\end{equation}

\subsection{Large Language Models}

Large Language Models are deep neural networks based on the Transformer architecture~\cite{vaswani2017attention}. The transformer architecture is composed of layers that each include a self-attention mechanism, allowing the model to relate different positions within the input sequence. This allows the model to compute relationships between all input tokens in a sequence. Self-attention enables the model to weigh the importance of different input tokens relative to each other. It does so by using $Q, K, V \in \mathbb{R}^{n \times d}$, where $n$ is the sequence length and $d$ is the embedding dimension. The matrices are constructed using learned linear projections of the input embeddings. In recent years, LLMs have been utilized for code generation tasks due to their ability to generate code from textual prompts with high performance. 

\subsection{Related Work}

Automatic regression test generation now combines classic symbolic-execution techniques with emerging LLM-based methods. Traditional methods remain influential, such as TracerX, which mitigates path explosion using interpolation and lazy annotations~\cite{jaffar2020tracerx}, or eXpress, which applies dynamic symbolic execution to concentrate on regression-relevant paths~\cite{taneja_express_2011}. Other symbolic execution–based tools include MutSyn for mutation-driven testing~\cite{su2023test} and map2check, which leverages program analysis for error detection and test generation~\cite{menezes2018map2check}.  

More recently, LLM-driven approaches have emerged. CoverUp integrates coverage feedback to guide test generation for Python programs~\cite{altmayer2025coverup}, while SymPrompt encodes execution constraints into prompts for systematic test creation~\cite{ryan2024codeawarepromptingstudycoverage}. Cleverest instead focuses on zero-shot prompt generation of failing regression tests, targeting structured input programs~\cite{liu2025llmgenerateregressiontests}.

\textit{UnitTenX} approaches the problem of unit test-generation by using formal verification to extract sensitization conditions to create unit that cause crashes. \textit{UnitTenX} is primarily aimed at legacy code bases with undocumented interfaces. By generating unit tests, we can resolve bugs and document unknown interfaces, thereby increasing their understanding.

\section{Motivating Example}
\label{sec:motivation}

\textit{AI Can Fix Bugs, But Can’t Find Them~\cite{venturebeat2025}:} 
Finding bugs corresponds to solving the reachability problem in sequential programs, which asks if an error state is attainable from initial conditions ~\cite{10298358}. Depending on variable domains and program complexity, this problem ranges from NP-complete implying high computational difficulty, to undecidable, where no algorithm guarantees a solution. Decoder-only LLMs generate sequences through a forward token generation process, and can at best enumerate solutions in token space, in an exponentially large search space that minimally requires backtracking~\cite{miserendino2025swe}. This limitation highlights the challenges faced by AI in identifying bugs, as the search space for potential solutions is vast and complex.

Consider the source code in Listing~\ref{lst:motivating_example}, which is an extracted function from \texttt{djbdns}~\cite{hansen_abhdjbdns_2025}, a DNS server implementation using the C programming language. In large software codebases, it can be challenging to ensure that the code behaves as expected, especially in scenarios where it relies on results from other independent parts of the software. In the example shown, it can be hard to assert with confidence that the code cannot enter a crash state. It can be just as hard to determine if the return statements in lines~\ref{lst:motivating_example:1} and~\ref{lst:motivating_example:2} are the only possible way to exit the function.

\begin{listingsbox}[label={lst:motivating_example}]{myc}{Function from a DNS server implementation}
int socket_recv4(int s, char *buf, int len, char ip[4], uint16 *port)
{
  struct sockaddr_in sa;
  int dummy = sizeof sa;
  int r;
  r = recvfrom(s, buf, len, 0, (struct sockaddr *)&sa, &dummy);
  if (r == -1) return -1;|\phantomsection\label{lst:motivating_example:1}|
  byte_copy(ip,4,(char *) &sa.sin_addr);
  uint16_unpack_big((char *) &sa.sin_port,port);
  return r;|\phantomsection\label{lst:motivating_example:2}|
}
\end{listingsbox}

To overcome this problem, we combine the use of LLMs with formal verification tools that can analyze the source code and provide a counterexample based on predefined properties that may have been violated~\cite{hasan2015formal, sanghavi2010formal, gadelha2018esbmc}. A counterexample is a description of the state of a program that has violated a given property~\cite{cordeiro2011smt, barreto2011verifying}. ESBMC~\footnote{\url{https://github.com/esbmc/esbmc}}~\cite{cordeiro2011smt}, a formal verification tool that automatically encodes safety properties such as integer arithmetic errors and buffer overflows, and checks for any possible violations, can be used to identify situations that may lead to crashes or other undesirable behavior~\cite{gadelha2018esbmc}, which is common in legacy software. We use ESBMC~\cite{cordeiro2011smt} to analyze C code for coverage and extract sensitization conditions. This approach complements the capabilities of LLMs, providing a more formal solution for bug detection and software testing.

Legacy code bases often lack comprehensive tests, making them difficult to maintain and modernize~\cite{feathers2004working}. Formal verification tools such as software model checkers can be used to find bugs and also to ensure the absence of bugs~\cite{biere2009handbook}. This often comes with its own set of challenges, for instance, with bounded model checkers, parameters such as the number of loop unwindings and execution timeouts must be carefully configured to balance the computational effort required to detect bugs against the necessary thoroughness and time of the search. These limitations motivate the need for automated, adaptive unit test generation systems that can efficiently improve test coverage and reliability in complex legacy systems.

\section{Methodology}
\label{sec:methodology}

\begin{figure}[ht]
    \centering
    \includegraphics[width=0.7\linewidth]{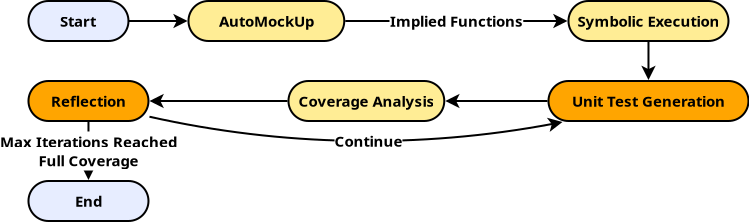}
    \caption{Shows each component of \textit{UnitTenX} along with the flow of data. The orange nodes denote steps that use LLM to process data.}
    \label{fig:unittenx_diag}
\end{figure}

\textit{UnitTenX} operates through a series of 5 steps, which are described in this section. A visual diagram is provided for reference in Figure~\ref{fig:unittenx_diag} and summarized below. The remainder of this section provides a detailed description of each step.

\begin{enumerate}
    \item \textit{AutoMockUps} generates mockups of each target function.
    \item \textit{Symbolic Analyzer} uses ESBMC, a formal verification tool to extract sensitization and crash conditions.
    \item \textit{Unit Test Generator} employs an LLM to produce unit tests.
    \item \textit{Coverage Analysis} compiles and reports coverage from the generated tests, using \texttt{gcov}.
    \item \textit{Reflection} uses the LLM to evaluate the results and recommends improvements to test quality and coverage.
\end{enumerate}

\begin{figure}[ht]
    \centering
    \includegraphics[width=0.48\linewidth]{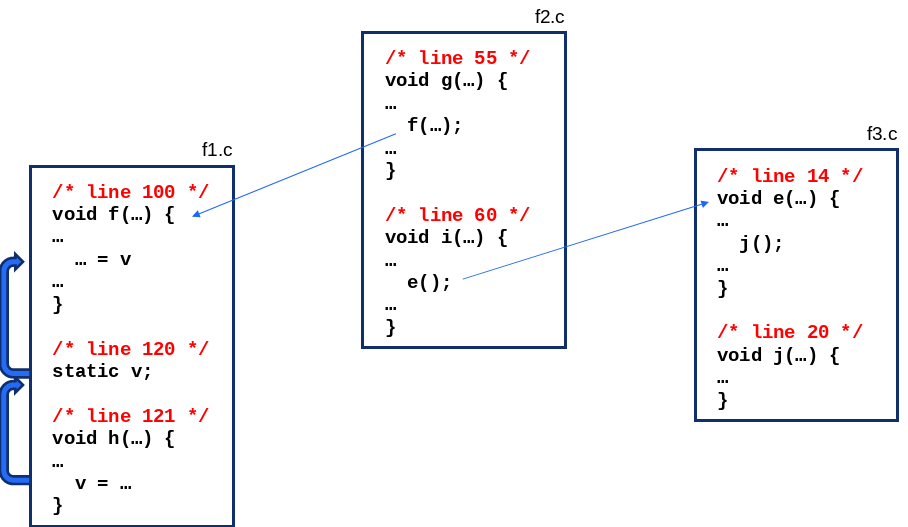}
    \caption{C Package Problems}
    \label{fig:c_package_problems}
\end{figure}

\textit{UnitTenX} processes code from a single source file when creating regression tests as the whole program is included in the context of the LLM. The \textit{AutoMockups} step automates this by constructing a single file for each function to be tested that contains the transitive closure of the symbol-dependency graph, called the \textit{Implied Functions}. The \textit{Implied Functions} are generated by iterating over all symbols in the source file, analyzing control-flow and dependency relationships between functions. This consolidation enables inclusion of the complete source context in LLM prompts for each target function, which is otherwise difficult because functions often depend on symbols across multiple files (Figure~\ref{fig:c_package_problems}). Additionally, because the automated process to generate the mockups is reversible, tests generated using formal methods and LLMs can be correlated and annotated in the original code.
By making functions and variables non-static, we increase the transparency and control over the test program, enabling more effective testing and analysis. Lastly, AutoMockups decreases long compilations when testing for coverage as it only compiles a single source file instead of multiple.

The \textit{Symbolic Analyzer} scans the single source file generated using AutoMockUps and extracts sensitization conditions that target coverage gaps and identifies potentially unsafe execution states. Using formal verification, it checks for vulnerabilities such as integer overflows/underflows, buffer overflows, and other crash-inducing behaviors that ESBMC automatically encodes~\cite{gadelha2018esbmc, biere2009handbook}. These extracted conditions then serve as input for the \textit{Unit-Test Generation} step.

\begin{figure}[ht]
    \begin{minipage}[b]{0.46\linewidth}
        \centering
        \includegraphics[width=\linewidth]{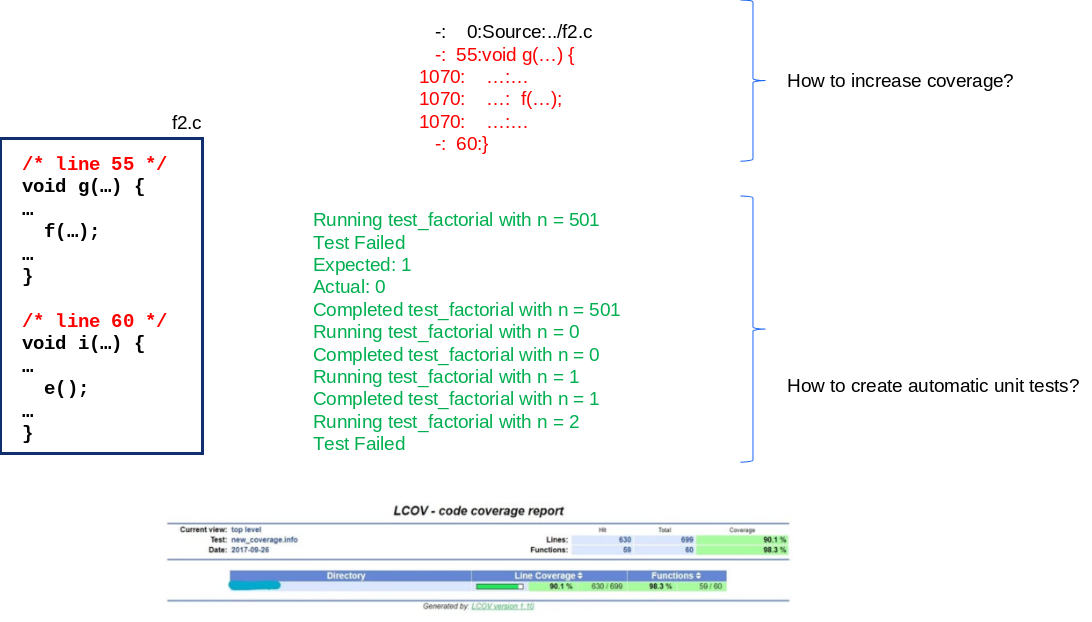}
        \caption{Tests are Regression Units}
        \label{fig:reg_units}
    \end{minipage}
    \begin{minipage}[b]{0.46\linewidth}
        \centering
        \includegraphics[width=\linewidth]{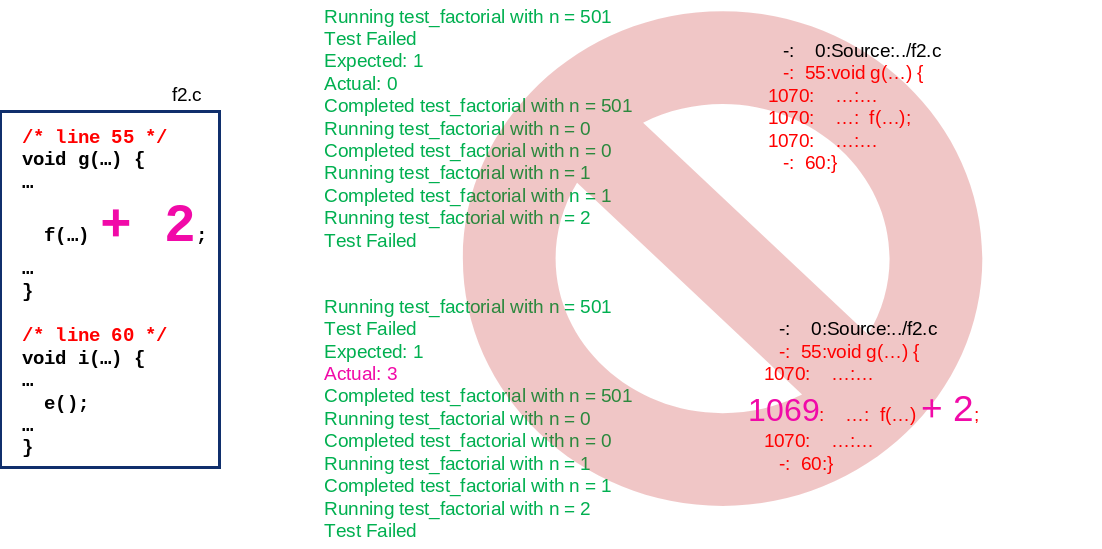}
        \caption{Regression Units Can Detect Changes}
        \label{fig:reg_units_detect_changes}
    \end{minipage}
\end{figure}

The Unit-Test Generation step creates unit tests using the LLM. The LLM receives in its input the entire \textit{Implied Function} C source code, previously generated tests, coverage analysis, and ESBMC outputs. The LLM then generates candidate unit tests. Any unit test that crashes on execution is commented out and annotated with \texttt{// CRASH}, creating a record of issues for developers and marking areas requiring further analysis. The unit-tests generated create a regression infrastructure, as illustrated in Figure~\ref{fig:reg_units}. This infrastructure allows developers to track changes in the code and assess their impact on the software’s behavior. Executing all generated tests can detect changes in package accesses or cross-dependencies, as illustrated in Figure~\ref{fig:reg_units_detect_changes}.

The \textit{Coverage Analysis} step compiles and extracts coverage information from the unit tests generated. This is later used in the reflection stage to evaluate the quality of the generated test. While ESBMC counterexamples are not used directly in this step, they influence coverage indirectly by guiding the LLM during unit test generation. If ESBMC completes with a decisive result, the unit-tests generated will usually have high coverage (as seen in Section~\ref{sec:eval}).

The Reflection step analyzes test and coverage results to recommend actions for improving tests. It is at this step that the test generation loop can exit or continue to keep improving the system. \textit{UnitTenX} will exit before the Reflection step if the max number of test generation and evaluation iterations has passed over a predefined value, and there are no errors reported from previous steps. If the loop is continued, the Reflection step tasks the LLM with rating the generated test by using the coverage results and recommending a plan of action for the \textit{Unit-Test Generation} step.

\section{Experimental Evaluation}
\label{sec:eval}

This section presents a comprehensive experimental evaluation of \textit{UnitTenX}~\footnote{\textit{UnitTenX} Source Code: \url{www.github.com/cnunescoelho/UnitTenX}}. A record of the experimental data and results can be found at~\cite{charalambous_2025_16642158}. The experiments were conducted using the legacy DNS server \texttt{djbdns}~\cite{hansen_abhdjbdns_2025} over 202 functions.
For the experimental execution, the tests involved the execution of \textit{UnitTenX} for each function, with automated logging to capture quantitative data for measuring code coverage, error handling, test generation, edge case analysis, and robustness. For the experiments, \textit{UnitTenX} utilizes the \texttt{pycparser} package for parsing C source code, ESBMC \texttt{v7.7.0 64-bit x86\_64} Linux for symbolic execution with Z3 \texttt{v4.13.3 64-bit} as the SMT backend. ESBMC is executed with a 10-second timeout, and GCC/clang with Gcov for instrumentation and coverage analysis. The LLM used to generate the unit tests and for review is \texttt{gpt-4o}~\cite{openai_gpt4o}. The max iterations of the \textit{Unit-Test Generation}, \textit{Coverage Analysis} and \textit{Reflection} steps was set to 4.

\subsection{RQ1: How effectively does \textit{UnitTenX} generate unit tests that increase code coverage for legacy C codebases?}

\begin{figure}[ht]
    \centering
    \begin{minipage}[c]{0.48\textwidth}
        \centering
        \includegraphics[width=0.8\linewidth]{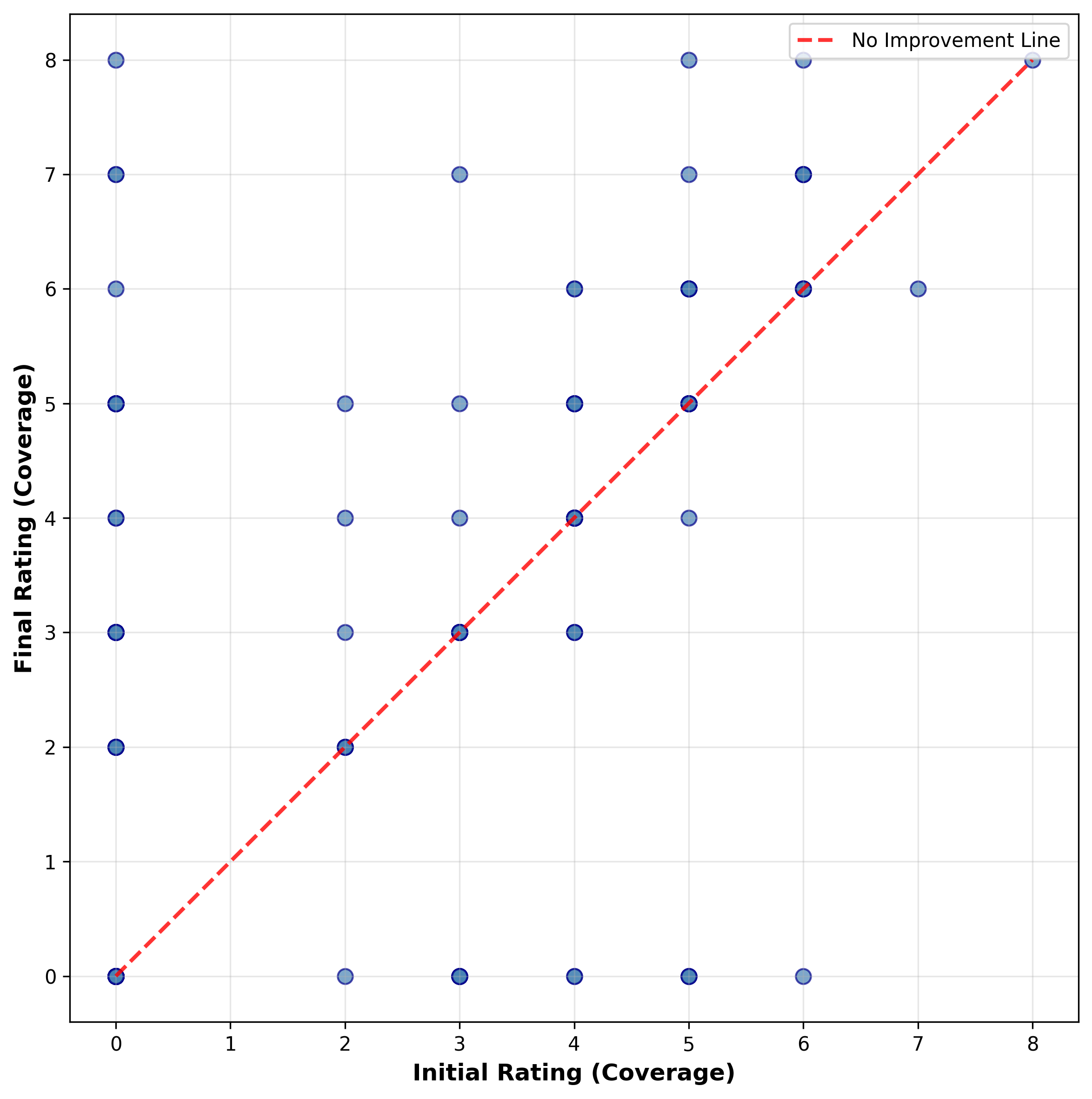}
        \caption{RQ1: \textit{UnitTenX} Initial test coverage\\rating against final.}
        \label{fig:rq1}
    \end{minipage}\hfill
    \begin{minipage}[c]{0.48\textwidth}
        \centering
        \includegraphics[width=0.8\linewidth]{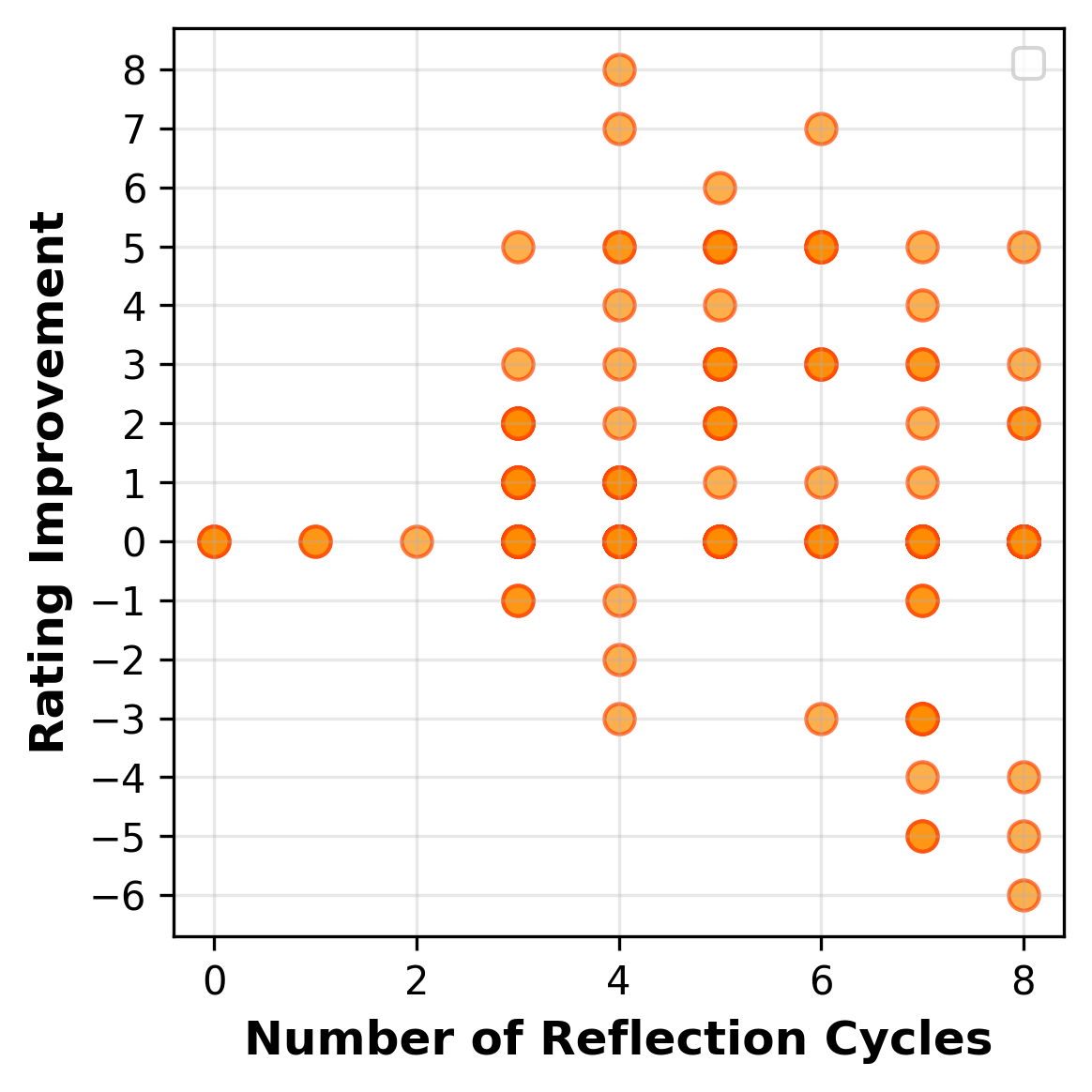}
        \captionof{figure}{RQ3: Reflection \& Feedback Loop Impact}
        \label{fig:rq3}
    \end{minipage}
\end{figure}

Figure~\ref{fig:rq1} illustrates a scatter plot comparing the initial test quality ratings, from the initial test generation (x-axis) against the final test quality ratings (y-axis) assigned by \textit{UnitTenX}'s reflection step. Each point represents a function, with its position indicating the initial and final subjective test quality ratings on a 0--8 scale, where points above the red dashed ``No Improvement Line'' signify improvement. Additionally, objective coverage metrics are derived from the dataset to assess coverage effectiveness from a baseline of 0\%, as no comprehensive unit tests existed before \textit{UnitTenX}'s intervention:

\begin{itemize}
    \item \textbf{Coverage Effectiveness:} Of the 199 functions that executed (98.5\% of 202 total), coverage was successfully measured for 186 functions (93.5\%).
    \item \textbf{Test Quality Improvement:} Figure~\ref{fig:rq1} shows that 66/199 functions (33.2\%) improved their test quality ratings, moving above the diagonal. The median test quality gain was +3 points (e.g., from 0 to 5), with the largest single improvement being from 0 to 8 (+8 points), as evidenced by the point at (0,8).
\end{itemize}

These findings highlight \textit{UnitTenX}'s dual capability: generating tests that cover previously untested code (186/199 functions) and enhancing test quality for a significant subset (33.2\%). While direct coverage percentages are not plotted, the high coverage success rate and test quality improvements show that \textit{UnitTenX} successfully generates comprehensive test suites for legacy C codebases.

\subsection{RQ2: How does \textit{UnitTenX} handle compilation errors, runtime exceptions (segmentation faults), and timeouts during automated test generation?}

\begin{itemize}
    \item \textbf{Compilation Errors:} The majority of failures occur due to code generated from the LLM that does not compile. There were 982 compilation errors. However, these were all resolved due to the iterative nature of \textit{UnitTenX}.
    \item \textbf{Segmentation Faults:} There were a total of 118 unit tests generated using the symbolic analyzer that crashed and were commented out. These expose crash conditions and are useful for documentation purposes.
    \item \textbf{Timeouts:} 61 total timeouts occurred when ESBMC exceeded its 10 second time limit.
\end{itemize}

Despite 1161 total errors across 1167 iterations, \textit{UnitTenX} produced compiling tests for all 199 executed functions, underscoring its robustness in recovering from intermediate setbacks.

\subsection{RQ3: How does the reflection and feedback loop in \textit{UnitTenX} contribute to iterative improvement of generated test suites?}

Figure~\ref{fig:rq3} shows a scatter plot of rating improvement (y-axis, -6 to 8) versus reflection cycles (x-axis, 0 to 8) for 199 executed functions. Each orange dot represents a function, with its position reflecting the change in test quality rating post-reflection. 66/199 functions (33.2\%) showed improved ratings. Pearson correlation yields $r = 0.0859$, $p = 0.224155$ ($p \geq 0.05$), indicating no significant relationship between reflection cycles and rating improvement. Most rating improvements occurred after the first three cycles, as can be seen by the shape of the diagram.

\subsection{Discussion}

The findings highlight \textit{UnitTenX}'s potential to significantly reduce the manual effort required for testing legacy codebases. In our evaluation, the codebase under test went from 0\% line coverage to 100\% proving that with the correct tools. A study showed that up to 50\% of the development time was spent on fault localization and bug fixing~\cite{rafi2023back}. By automating the generation of high-coverage test suites and effectively handling errors, \textit{UnitTenX} addresses key challenges in maintaining and modernizing legacy software systems. The ability to generate tests that expose crash conditions also enhances the documentation and understanding of legacy interfaces, which is critical for long-term maintenance. Moreover, the system's robustness in error handling makes it a reliable tool for production environments.

\section{Conclusion}
\label{sec:conclusion}

\paragraph{Threats to Validity} Several limitations should be considered when interpreting the results. First, the evaluation was conducted on a single legacy DNS server code-base. This limits the generalizability of the results. Second, the test quality ratings used in the reflection step are subjective (to the LLM's generation) and could introduce bias in assessing test quality improvement. Objective metrics, such as code coverage percentages or fault detection rates, would provide a more rigorous evaluation.

This paper introduced \textit{UnitTenX}, a tool to generate regression tests to document legacy codebases automatically by using LLMs equipped with formal verification to extract sensitization conditions. The experimental evaluation showed that \textit{UnitTenX} was able to bring 100\% line coverage to a real-life code-base without any tests initially, proving that it is capable of being used in production environments. This makes it a viable tool for software maintenance and modernization. However, further research is needed to validate its effectiveness across diverse legacy software systems.

To address these limitations, future research should evaluate \textit{UnitTenX} on a broader range of legacy codebases, including those from different domains and with varying levels of complexity. Integrating more objective metrics, such as coverage into the rating process in the \textit{Reflection} step could improve its overall accuracy in assessing the generated unit-tests.

\paragraph{Declaration on Generative AI} During the preparation of this work, the author(s) used Perplexity AI to review documents and sentences. After using these tool(s)/service(s), the author(s) reviewed and edited the content as needed and take(s) full responsibility for the publication’s content.

\bibliography{references}

\end{document}